\documentclass[twocolumn,showpacs,preprintnumbers,amsmath,amssymb]{revtex4}

\usepackage{graphicx}
\usepackage{dcolumn}
\usepackage{bm}
\graphicspath{{./}{./figs/}}

\begin{document}

\title{\bf Stability of a neural network model with small-world connections}
\author{Chunguang Li}
\email{cgli@uestc.edu.cn}
\affiliation{Institute of Electronic Systems, College of Electronic Engineering,\\
University of Electronic Science and Technology of China,\\
Chengdu, Sichuan, 610054, P. R. China.}
\author{Guanrong Chen}
\email{gchen@ee.cityu.edu.hk}
\affiliation{Department of Electronic Engineering, City University of Hong Kong, \\
83 Tat Chee Avenue, Kowloon, Hong Kong, P. R. China.}

\date{\today}
\begin{abstract}
Small-world networks are highly clustered networks with small
distances among the nodes. There are many biological neural
networks that present this kind of connections. There are no
special weightings in the connections of most existing small-world
network models. However, this kind of simply-connected models
cannot characterize biological neural networks, in which there are
different weights in synaptic connections. In this paper, we
present a neural network model with weighted small-world
connections, and further investigate the stability of this model.
\end{abstract}
\pacs {89.75.Hc, 87.18.Sn, 84.35.+i, 05.45.-a,}
\maketitle

A great deal of research interest in the theory and applications
of small-world networks have arisen [1-8] since the pioneering
work of Watts and Strogatz [9]. Some common properties of complex
networks, such as Internet servers, power grids, human
communities, and disordered porous media, are mainly determined by
the way of connections among their vertices or nodes. Among
various networks, one extremal case is a regular network with a
high degree of local clustering and a large average distance,
while the other extremal case is a random network with negligible
local clustering and a small average distance. In between the two
extremes there are small-world networks, which are a special type
of complex networks with a high degree of local clustering as well
as a small average distance.

Many biological neural networks are small-world networks [10-13].
In most existing literature about small-world networks, there are
no weightings in their internal connections of nodes. However,
there are many networks, particularly biological neural networks,
having weights associated with the connections. These
connection-weighted networks cannot be described and characterized
by those previously proposed small-world network models. Of
particular interest is [13], where small-world neural networks
have random weights in their connections, which studied the
cluster coefficient and the characteristic path of such networks.
In this paper, we use dynamical equations to describe a
connection-weighted small-world neural network model, and then
further study its stability with respect to the network topology.

For this purpose, consider a neural network with $N$ neurons
described by
\begin{equation}
\frac{\mbox{d}u(t)}{\mbox{d}t}=-Au(t)+Wg(u(t))+I
\end{equation}
where $u(t)=\left[u_1(t),u_2(t),\cdots,u_N(t)\right]^T$ is the
neuron state vector, $A=\mbox{diag}\{a_1,a_2,\cdots,a_N\}$ is a
positive diagonal matrix, $g(u)=\left[g_1(u_1),g_2(u_2),\cdots,
g_N(u_N)\right]^T$ denotes the neuron activation functions with
$g(0)=0$, $I=[I_1,I_2,\cdots,I_N]^T$ is a constant vector,
$W=\{w_{ij}\}_{N\times N}$ is the connection-weighting matrix, in
which, similar to [13], $w_{ij}$ is defined as follows: if there
is a connection between neuron $i$ and neuron $j$ $(j\neq i)$,
then there is a uniform random distribution $w_{ij}=w_{ji}$ in the
connection, with values $0<w_{ij}=w_{ji}< 1$; otherwise,
$w_{ij}=w_{ji}=0 \,(j\neq i)$. The diagonal elements of $W$ are
all zeros, which means that there are no self-connection of nodes
within the network. Throughout this paper, we assume that each
activation function in (1) satisfies the following sector
condition: There is a real constant, $k \in R$, such that
\begin{equation*}
0\leq\frac{g_j(x)-g_j(y)}{x-y}\leq k,\hspace{0.2cm}\forall\,x,y
\in R,\hspace{0.2cm} j=1,2,\cdots,N
\end{equation*}

This type of neural networks with full and regular connections has
been extensively investigated. However, neural networks with
small-world connections have not been thoroughly studied,
particularly with respect to their stabilities. For example, it is
not clear whether or not the small-world neural networks are
easier to be stabilized than the fully connected ones. In this
paper, we address this question by carefully studying model (1).

In the following, we always shift the equilibrium $u^*$ of network
(1) to the origin. By making the transform $x(t)=u(t)-u^*$, we
convert model (1) to the following:
\begin{equation}
\frac{\mbox{d}x(t)}{\mbox{d}t}=-Ax(t)+Wf(x(t))
\end{equation}
where $f_j(x_j(t))=g_j(x_j(t)+u_j^*)-g_j(u_j^*),\,j=1,2,\cdots,N$.
Note that $f_j$ also satisfies a sector condition in the form of
\begin{equation}
f_j(x_j)\left(f_j(x_j)-kx_j\right)\leq 0,\hspace{0.5cm}
j=1,2,\cdots,N
\end{equation}

For this small-world neural network model, we have the following
theoretical result: \smallskip

{\it Lemma}: Let $\lambda_{max}(M)$ denote the largest eigenvalue
of matrix $M$. If $\lambda_{max}\left(-\frac{A}{k}+W\right)<0$,
then network (2) is asymptotically stable about the origin.
\smallskip

{\it Proof}: Select a Lyapunov function as
\begin{equation*}
V(x(t))=\sum_{i=1}^N\int_0^{x_i}f_i(s)ds
\end{equation*}
Using the method presented in [14, 15], it is easy to verify that
$V(x(t))$ is a Lyapunov function. In fact, if we define the
following function:
\begin{equation*}
G_i(u)=\mbox{min}\left\{\int_0^uf_i(\theta)d\theta,
\int_0^{-u}f_i(\theta)d\theta\right\}
\end{equation*}
Then we have $G_i(0)=0, G_i(u)=G_i(|u|)$; for $r\in R^+$,
$G_i(r)>0, r>0; G_i(r)\rightarrow +\infty$, $r \rightarrow
+\infty$. Let $G=\mbox{min}\{G_i\}$. Then
\begin{equation*}
\begin{array}{rl}
V(x)&=\sum_{i=1}^N\int_0^{x_i}f_i(s)ds\\
&\geq \sum_{i=1}^N G_i(x_i)\\
&=\sum_{i=1}^N G_i(|x_i|)\\
&\geq \sum_{i=1}^N G(|x_i|)\\
&\geq G(|x|)
\end{array}
\end{equation*}
Therefore, we have a lower bound achieved by a positive, radially
unbounded function. It is then easy to verify [15] that
\begin{equation*}
G(|x(t)|)\leq V(|x(t)|)\leq qk\|x\|^2,\,q>1
\end{equation*}

The derivative of $V(x)$ along the trajectories of (2) is, by
using (3),
\begin{equation*}
\begin{array}{rl}
\dot{V}(t)=&\left[f_1(x_1),\cdots,f_N(x_N)\right]
\left[\dot{x}_1,\cdots,\dot{x}_N\right]^T\\
=&f^T(x(t))\left[-Ax(t)+Wf(x(t))\right]\\
=&-f^T(x(t))Ax(t)+f^T(x(t))Wf(x(t))\\
\leq&f^T(x(t))\left[-\frac{A}{k}+W\right]f(x(t))\\
\leq&\lambda_{max}\left(-\frac{A}{k}+W\right)\|f(x(t))\|^2
\end{array}
\end{equation*}
Therefore, if $\lambda_{max}\left(-\frac{A}{k}+W\right)<0$, then
we have $\dot{V}(x(t))<0$, implying that network (2) is
asymptotically stable about the origin. \smallskip

Because $A$ is a diagonal positive matrix, it is easy to deduce
the following Corollary. \smallskip

{\it Corollary}: If $\lambda_{max}(W)<\mbox{min}\{\frac{a_i}
{k}\}$, then network (2) is asymptotically stable about the
origin. \smallskip

Although we can derive some less conservative stability conditions
for (2), we only use the above results in this paper, because they
are very simple and easy to verify. Further, because these
conditions use only the maximum eigenvalue of the connection
matrix $W$, we can ``average'' them when using statistical methods
to investigate the properties of the connection matrix, as further
explained in the following.

Aiming to describe a transition from a regular network to a random
network, [9] introduced an interesting model, now referred to as
the small-world (SW) network. The original SW model can be
described as follows. Take a one-dimensional lattice of $N$
vertices arranged in a ring with connections only in between
nearest neighbors. We then ``rewire'' each connection with
probability $p$. Rewiring in this context means reconnecting
randomly the whole lattice, with the constraint that no two
different vertices can have more than one connection in between,
and no vertex can have a connection with itself.

Note, however, that it is quite possible for the SW model to be
broken into unconnected clusters. This problem can be resolved by
a slight modification of the SW model, suggested by Newman and
Watts (NW) lately [1]. In the NW model, we do not break any
connection between any two nearest neighbors. Instead, we add with
probability $p$ a connection between each unconnected pair of
vertices. Likewise, we do not allow a vertex to be coupled to
another vertex more than once, or a vertex to be coupled with
itself. For $p=0$, it reduces to the originally nearest-neighbor
coupled network; for $p=1$, it becomes a globally coupled network.
Here, we are interested in the NW model starting from a
nearest-neighbor lattice with 4-neighbors and a connection-adding
probability $0<p<1$.

 From a coupling-matrix point of view, network (2) with small-world
connections evolves according to the rule that, in the
nearest-neighbor coupling matrix $W$, if $w_{ij}=0$, we set
$w_{ij}=w_{ji}=w$ with probability $p$ and a uniformly randomly
distributed weight $0<w<1$. We denote the new small-world coupling
matrix by $W(p,N)$ and let $\lambda_{max}(p,N)$ be its largest
eigenvalue. According to Corollary 1, if $\lambda_{max}(p,N)<
\mbox{min}\{\frac{a_i}{k}\}$, then the corresponding small-world
neural network is asymptotically stable about its zero state.

Clearly, the network stability depends on the probability $p$, so
it is more practical to investigate the statistical properties of
the connection matrix $W$. It is easy to see that the mathematical
expectation of the number of neurons that are connected to each
neuron, i.e., the number of nonzero entries in each row of
$W(p,N)$, is $n_c=4+(N-5)p$. Because of the uniform random
distribution of the weight values, it is also easy to see that the
mathematical expectation of the sum of entries in each row of
$W(p,N)$ is $0.5[4+(N-5)p]$, where $N\geq 5$ (the smallest neuron
number of the nearest-neighbor lattice with 4-neighbors). Thus, by
Lemma 2 of [16], we can calculate the mathematical expectation of
$\lambda_{max}(p,N)$, which is $0.5[4+(N-5)p]$, where $N\geq 5$.
Hence, the small-world neural network is asymptotically stable
about its zero state, in the sense of mathematical expectation, if
$0.5[4+(N-5)p]<\mbox{min}\{\frac{a_i}{k}\}$, $N\geq 5$. This also
means that the small-world neural network is asymptotically stable
in the sense of mathematical expectation if the number of
connections of each neuron $n_c$ in the networks is
$n_c<2\mbox{min}\{\frac{a_i}{k}\}$.

\begin{figure}[htb]
\centering
\includegraphics[width=6cm]{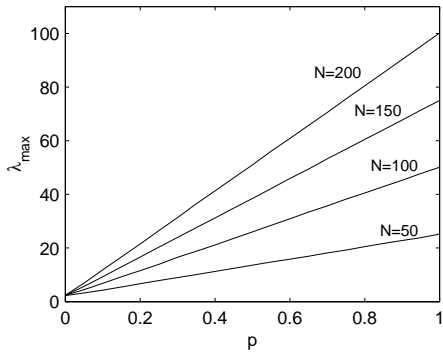}
\caption{Maximum eigenvalue of $W$ w.r.t. $p$}
\end{figure}
\begin{figure}[htb]
\centering
\includegraphics[width=6cm]{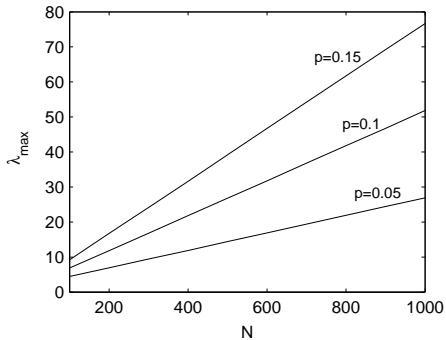}
\caption{Maximum eigenvalue of $W$ w.r.t. $N$}
\end{figure}

Figures 1 and 2 show the numerical values of $\lambda_{max}(p,N)$
as a function of the probability $p$ and the number of neurons
$N$. In these figures, for each pair of values $p$ and $N$,
$\lambda_{max}(p,N)$ is obtained by averaging the results of 20
runs. From the above analysis and the above figures, we can see
that
\begin{enumerate}
 \item for any given value of $N\geq 5$, $\lambda_{max}(p,N)$
increases almost linearly from about 2 to about $(N-1)/2$ as $p$
increases from 0 to 1;
 \item for any given value of $p\in (0,1]$, $\lambda_{max}(p,N)$
increases almost linearly to $+\infty$ as $N$ increases to
$+\infty$.
\end{enumerate}

The above results imply that if the given matrix $A$ satisfies
min$\{\frac{a_i}{k}\}>2$ then
\begin{enumerate}
\item for any given $N\geq 5$, there exists a critical value
$p^*$ such that if $0\leq p \leq p^*$, then the small-world neural
network is asymptotically stable about its zero state (in the
sense of mathematical expectation);
 \item for any given $p\in(0,1]$, there exists a critical value
$N^*$ such that if $5\leq N\leq N^*$, then the small-world neural
network is asymptotically stable about its zero state (in the
sense of mathematical expectation).
\end{enumerate}

Clearly, neural networks with small-world connections are easier
to be stabilized than their regular fully-connected counterparts.
\smallskip

Next, we consider an example of network (1), with the constant
vector $I=0$, $A=\mbox{diag}\{5,5,\cdots,5\}$, and the activation
condition $g(\cdot)$=tanh$(\cdot)$, which also satisfies condition
(3).

 From the above results, we know that for any given $N$ (or any
given $p$), there exist a corresponding $p$ (or a corresponding
$N$) that guarantee the stability of the network. The shadow zone
in Figure 3 shows the values of $p$ and $N$ that ensure the
stability of this small-world neural network. In this example, the
averaged number of connections of each neuron in various
configurations is $n_c=9.8592$. This result also coincides with
the above analysis.

\begin{figure}[htb]
\centering
\includegraphics[width=6cm]{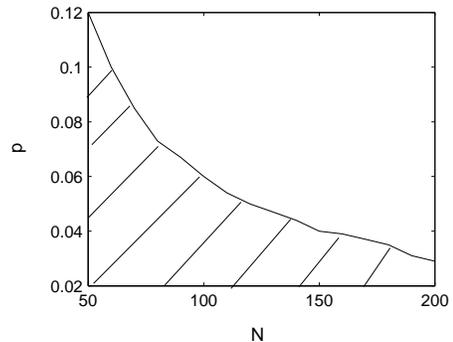}
\caption{Stability zone of the network in the given example}
\end{figure}

In summary, a small-world neural network model has been presented
and its stability has been analyzed. An analytical expression,
which establishes the relationship between the stability and the
probability $p$, has been derived. Because there are many
biological neural networks that present small-world connections,
the results obtained in this paper are practical and should be
useful for further studies of this kind of neural network models.
\smallskip

We are grateful to the anonymous reviewers for their valuable
comments and suggestions. We acknowledge supports from the
National Natural Science Foundation of China under Grant 60271019
and the Hong Kong Research Grants Council under the CERG grant
CityU 1004/02E.

\end{document}